\documentclass[12pt]{iopart}

\usepackage{amsfonts}
\usepackage{booktabs}
\usepackage{cite}
\usepackage{graphicx}
\usepackage{float}
\usepackage{subfig}
\usepackage{color}
\usepackage{bm}

\begin{document}

\title[Topology optimized ERT based sensors]
{Topology optimization enhances the distinguishability and reconstructability of electrical resistance tomography based sensors}

\author{Reza Rashetnia$^1$ and Mohammad Pour-Ghaz$^2$*}
\address{$^1$ Knowbe4 Inc., USA}
\address{$^2$ Department of Civil Construction and Environmental Engineering, North Carolina State University, Raleigh, NC 27695, USA}
\ead{reza.rashetnia@gmail.com}

\begin{abstract}
In the majority of applications of electrical resistance tomography (ERT) the estimation problem consists of either the estimation of spatial conductivity change over an existing background or the estimation of spatial distribution of conductivity of the entire target, including the background.
In some instances however, it is possible to design the background conductivity; an example of such application is the design of ERT-based sensors where the background conductivity can be engineered. 
In such applications the natural question is whether the background conductivity can be engineered in such a way to increase the distinguishability and further reconstructability of the sensor. 
The present paper, uses topology optimization to design the background conductivity to achieve optimal distinguishability.
Then, ERT reconstructions suggest the enhancements of reconstructability using topology optimized sensor.  

\end{abstract}
\maketitle

\section{Introduction}
 \label{INT}

Electrical Resistance Tomography (ERT) is an imaging modality in which the spatial distribution of electrical conductivity (or resistivity) of a target is reconstructed based on a set of electric current injections and their corresponding measured potentials \cite{holder2005electrical}. 
For this purpose, typically electrodes are installed at the boundary of a target; the same set of electrodes are used for current injections and potential measurements.

ERT has many medical and industrial applications. 
In the majority of these applications, the goal is to estimate the change of conductivity over an existing background \cite{Brown2003,Bodenstein2009,Costa2009,Hallaji2014,Rashetnia2018,Rashetnia20182}, to estimate the spatial distribution of conductivity of the entire target including the background\cite{Cheney1999,Brown2003,Cherepenin2002,Hallaji2014b,Rashetnia2017}, or simultaneously estimate a change and the background spatial conductivity distribution \cite{Liu2015,Liu2015b}. 
In these applications the background conductivity may be unknown, non-uniform, or otherwise. 
In some instances however, there is a possibility of designing the background conductivity to achieve a certain objective. 
An example of such application is the so-called ERT-based sensing skin (referred to as sensing skin hereafter)\cite{Hallaji2014b,Rashetnia2017,Hallaji2014,Rashetnia2018,Rashetnia20182}.

Sensing skin is a thin layer of electrically conductive materials that is applied to the surface of a structure in the form of paint or prefabricated wallpaper\cite{Hallaji2014}. 
Any change of conductivity, resulting from a stimulus such as damage, is detected and quantified using ERT. 
In this particular application, the background conductivity (the initial conductivity distribution over the sensing skin) can be designed during the manufacturing of the sensor to achieve a certain objective \cite{Rashetnia}. 
For example, background conductivity can be designed to achieve higher "detection resolution" away from the boundaries of the sensing skin where the measurements are performed. 
While the present paper, uses example of sensing skin, the problem solved herein is more general: Can the background conductivity be designed to achieve a higher distinguishability over the entire domain? 

In the present work, we use topology optimization \cite{Bends,Sigmund1994} to optimize the background conductivity to yield a higher average distinguishability over the entire domain. 
Herein the distinguishability defined by Issacson \cite{Isaacson1986} (as opposed to resolution) is used to optimize the background conductivity since a rigorous definition for resolution does not exist due to the ill-posed nature of ERT reconstruction; we note that a higher distinguishability may not necessary mean a better reconstructability due to the ill-posed nature of the ERT inverse problem.
Therefore, ERT inverse problem is performed to study whether distinguishability enhancement can provide a better reconstructability.

We also note that optimization for distinguishability is dependent on the geometry of the target, the current injection and potential measurement pattern, as well as the shape and orientation of the anomaly to be detected. 
To simplify the problem therefore, we use a two-dimensional circular domain, a circular anomaly, and symmetric current injection and potential measurements. 
While the results obtained in the current work are only applicable to the problem solved herein, the methodology is general and can be applied to other targets and anomalies of interest.

\section{ERT domains description}
\label{ERT1}

To simulate ERT experiments, a 2D domains is computationally simulated in this paper which is shown in Figure 1a.
For simplicity, a symmetric circular ERT domain ($\Omega$) with radius of 1 is considered.
Eight equally spaced electrodes are positioned over the boundary ($\partial\Omega$).
The length of electrodes are kept constant as $0.2$.
The electrical conductivity distribution is constrained between $0.01 \leq \sigma \leq 1$ (non-zero to avoid singularity).
ERT measurements are applied on the domain with all current injection and potential measurement patterns possible.
Current injection is considered as $I_{ij}$ where current of $1$ amp injected from electrode $i$ and ejected form electrode $j$; $I_{ij} =  \{ I = 1 |  i,j =1 , ... , e \}$, where $e$ is number of electrodes.
Electrical potentials are measured at all electrodes for each current stimulations; $U_{ij}^{k},  k =1 , ... , e$ , where $k$ stands for electrodes.

\begin{figure*}[ht!]
\centering
\label{pa}
\subfloat[]{\includegraphics[width=4.5cm]{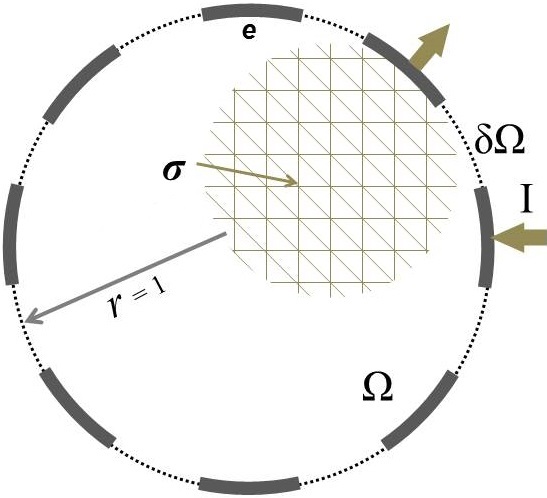}}%
\qquad
\subfloat[]{\includegraphics[width=4.5cm]{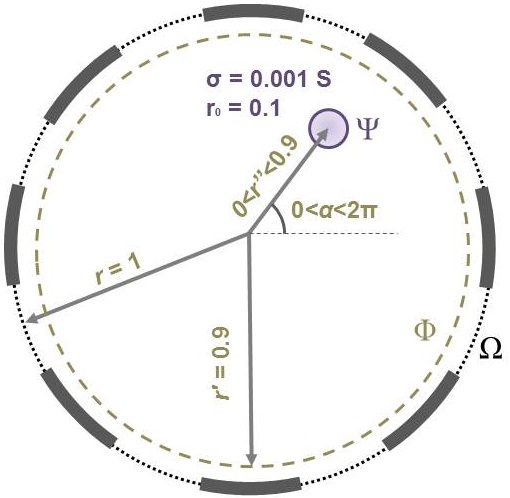}}%
\caption{ a) Description of a symmetric circular domains, $\Omega$, with eight electrodes over its boundaries, $\partial \Omega$ ($0.01 \leq\sigma \leq 1$); b) The anomaly $\Psi$ ($\Psi \subset \Phi \subset \Omega$) is moved inside $\Phi$.}
\end{figure*}

\section{ERT complete electrode model} 
\label{FWD}

The Complete Electrode Model (CEM) is used which is the most accurate forward model for the ERT since it takes into account the effects of the electrodes and contact impedances \cite{cheng1989electrode,somersalo1992}.
The CEM consists of the differential equation

\begin{equation}
\label{CEM}
   \nabla \cdot ( \sigma \nabla u ) = 0, \ \ \ x\in\Omega,
\end{equation}

\noindent 
and
boundary conditions

\begin{equation}
\label{BC1}
   u + \xi_{i} \sigma \frac{du}{d\bar{n}} = U_{i}, \ \ \ x \in e_{i},\ i=1,\ldots,L
\end{equation}

\begin{equation}
\label{BC3}
   \int_{e_{i}}\sigma \frac{du}{d\bar{n}} dS = I_{i}, \ \ \ i=1,\ldots,L
\end{equation}

\begin{equation}
\label{BC2}
   \sigma \frac{du}{d\bar{n}} = 0, \ \ \ x\in \partial\Omega \backslash \bigcup_{i=1}^{L} \varrho_{i}
\end{equation}

\noindent
where $\sigma$ is the electrical conductivity, $u$ is the electric potential, $\bar{n}$ is the outward unit normal, \begin{math}{\varrho_i}\end{math} is the surface area under \begin{math}{i^{\mathrm{th}}} \end{math} electrode, and
\begin{math} {\xi_i}\end{math}, ${U_i}$ and ${I_i}$,  respectively, are the contact impedance, electric potential and total current corresponding to \begin{math}{\varrho_i}\end{math}.
Further, the charge conservation law must be satisfied, and the potential reference level needs to be fixed,
\begin{equation}
\label{BC4}
   \sum_{l=1}^L I_l = 0, \  \
    \sum_{l=1}^L U_l = 0.
\end{equation}
\noindent

Here, Galerkin Finite Element Method (GFEM) with piecewise linear approximation of electrical conductivity and quadratic approximation for electrical potential is used to approximate the solution of the variational form of the forward model \cite{Vauhkonen1999,Vauhkonen2001,vauhkonen04}.
Using GFEM, the potential distribution within the object is approximated with the finite sum

\begin{equation}
\label{FE1}
   u^{h}(x) = \sum_{i=1}^{N'} \alpha_{i}\phi_{i}(x)
\end{equation}
\noindent
and the potentials on the electrodes as 

\begin{equation}
\label{FE2}
   U^{h}(x) = \sum_{j=1}^{L-1} \beta_{j} n_{j}
\end{equation}

\noindent
where the functions $\phi$ and $n_{j}$ are the basis functions and $\alpha$ and $\beta$ are the nodal potential values of $\Omega$ and $\partial\Omega$. 
In Equation \ref{FE1}, $N'$ is  the number of nodes.
Inserting these two approximative functions into the variational equations results in a system of linear equations which can be written in matrix form as

\begin{equation}
\label{FE3}
   \textbf{K} \textrm{V}= \textrm{f}
\end{equation}

\noindent
where $\textrm{V}=(\alpha,\beta)^{T}$ are the electrical potentials. $\textrm{f}=(0,\hat{I})^{T}$, where $\textrm{0} \in \mathbb{R}^{1\times N'}$ and $\hat{I}$ is the vector of current of current stimulation pattern.
$\textbf{K}$ is the stiffness matrix which takes into account the effects of the electrodes and contact impedances between the object and the electrodes \cite{cheng1989electrode,somersalo1992}. Thus the approximation for the potentials $V$ are obtained by solving Equation \ref{FE4}

\begin{equation}
\label{FE4}
   \textrm{V}= \textbf{K}^{-1}  \textrm{f}.
\end{equation}

\section{Distinguishability}
 \label{DIS}

Two conductivities $\bm{\sigma}_{1}$ and $\bm{\sigma}_{2}$ are distinguishable when the difference between their potential measurements exceeds the experimental percision of measurements, $\epsilon$ \cite{Isaacson1986,Cheney1992,garde2016depth}.
Equation \ref{dis} presents the distinguishability criterion, $\Lambda$.

\begin{equation}
\label{dis}
 \Lambda(\bm{\sigma}_{2},\bm{\sigma}_{1},\textbf{I}) = \frac{\| \textbf{U}(\bm{\sigma}_{2},\textbf{I}) - \textbf{U}(\bm{\sigma}_{1},\textbf{I})\|}{\|\textbf{U}(\bm{\sigma}_{1},\textbf{I}).\textbf{I}\|} > \epsilon
\end{equation}

In \cite{Isaacson1986,Cheney1992,garde2016depth}, distinguishability is defined also as "homogeneous of degree zero" to be the norm of the voltage measurement differences over electrodes divided by the norm of the power applied to $\bm{\sigma}_{1}$.
Therefore, distinguishability criterion remains neutral to the amplitude of current injection and projects the potential changes between two states of electrical conductivities. 
According to this definition, the most desirable ERT measurements have higher distinguishability.

The distinguishability distribution over $\Omega$, can be estimated using Equation \ref{dis} between two conductivity distribution of $\bm{\sigma}_{1}$ and $\bm{\sigma}_{2}$.
$\bm{\sigma}_{1}$ refers to the original domain, and $\bm{\sigma}_{2}$ refers to the same domain that an anomaly with different conductivity is placed inside of it.
Here, a circular anomaly inside the domain $\Psi \subset \Omega$ with radius of $0.1$ and electrical conductivity of 0.001 $\ll$ 1  is moved inside $\Phi$ over the entire domain (Figure 1b).
The spatial distribution of distinguishability, $\Lambda$, then is estimated over $\Phi$.

In this paper, we aim to maximize $\Lambda$ with respect to the optimal background electrical conductivity distribution of the domain.
$\Lambda(\bm{\sigma}_{2},\bm{\sigma}_{1},\textbf{I})$ is maximized by maximizing $\textbf{U}(\bm{\sigma},\textbf{I})$.
$\textbf{U}(\bm{\sigma},\textbf{I})$ is maximized when the electrical power transferred through electrodes are maximized  (Equation \ref{optn}).
Therefore, we maximize the energy transferred in the domain based on background conductivity distribution using a maximization algorithm of compliance matrix over the domain to optimize background conductivity distribution with constrained conductivity and subjected to the CEM forward model.

\begin{equation}
\label{optn}
\widehat{\Lambda} = \arg\max_{\bm{\sigma}_{1}} \Lambda(\bm{\sigma}_{2},\bm{\sigma}_{1},\textbf{I}) 
\end{equation}

\section{Optimization Problem}
 \label{TOP}

Equation \ref{comp} defines the compliance matrix, $\textbf{P}(\bm{\sigma})$ which is discretized FEM electrical power over the discretized domain, $\Omega$,

\begin{equation}
\label{comp}
\textbf{P}(\bm{\sigma}) = \textbf{V}^{T}\textbf{K}\textbf{V}.
\end{equation}

In this section,  the power-law approach is used to maximize $\textbf{P}(\bm{\sigma})$ as a function of $\bm{\sigma}$ in the domain. 

Equation \ref{opt1} shows the optimization of $\textbf{P}(\bm{\sigma})$ resulting in $\widehat{\bm{\sigma}}$

\begin{equation}
\label{opt1}
\widehat{\bm{\sigma}} = \arg\max_{\bm{\chi}} \Big\{ \| \textbf{V}^{T}\textbf{K}\textbf{V}\|_{1/2}\Big\} = \arg\max_{\bm{\chi}} \Bigg\{ \sum_{i=1}^{N} (\chi_{i})^{\kappa} v_{i}^{T}k_{i} v_{i} \Bigg\} \nonumber
\end{equation}
\noindent
which is subjected to: 
\begin {eqnarray}
\frac{\bm{\phi}(\sigma)}{\bm{\phi}_{0}}=\zeta 
\\
\bm{\sigma} = \bm{\chi} \otimes \bm{\sigma}_{0} \nonumber
\\
\textbf{K} \textrm{V}= \textrm{f} \nonumber
\\
0.01 \leq \bm{\sigma} \leq 1 \nonumber
\end{eqnarray}

\noindent
where $\textbf{K}$ is global stiffness matrix, $v_{i}$ and  $k_{i}$ are the elements of potential vector and stiffness matrix, $\bm{\chi}$ is the vector of design variable for electrical conductivity, and $\bm{\sigma}$ is tensor of electrical conductivities, 
$N$ is the number of elements and $\kappa$ is penalization power which is normally 3 \cite{Bends,Sigmund1994}.
$\bm{\sigma}_{0}$ is unit tensor of electrical conductivities, and $\bm{\phi}(\sigma)$ and $\bm{\phi}_{0}$ are designing target and initial assembled electrical conductivity respectively.
$\zeta$ is the constraint condition for volume fraction of electrical conductivity distributions.
The optimization problem (Equation \ref{opt1}) is solved using standard Optimality Criteria method \cite{Bends,Sigmund1994}.

A heuristic updating method for electrical conductivity is used based on Bends${\o}$e \cite{Bends} which is formulated as

\begin{equation}
\label{opt2}
\sigma_{i}^{new} = \left\{
\begin{array}{ll}
 \textrm{if}:  \chi_{i} \rho_{i}^{\mu} \leq \max (\sigma_{\min},\sigma_{i}-m)
\\
\quad \quad \max(\sigma_{\min},\sigma_{i}-m) \quad 
\\
\textrm{if}: \max(\sigma_{\min},\sigma_{i}-m) < \chi_{i} \rho_{i}^{\mu} < \min (1,\sigma_{i}+m)
\\
\quad \quad \chi_{i} \rho_{i}^{\mu} \quad 
\\
\textrm{if}: \min (1,\sigma_{i}+m) \leq \chi_{i} \rho_{i}^{\mu},
\\
\quad \quad \min (1,\sigma_{i}+m) \quad 
\end{array}
\right.
\end{equation}

\noindent
where $m$ is incremental move limit, $\mu = 1 / 2$ is a numerical damping coefficient and $\rho_{i}$ is found from the optimality condition as

\begin{equation}
\label{opt3}
   \rho_{i} = \frac{-\frac{\partial p}{\partial \chi_{i}}}{\lambda \frac{S}{\partial \chi_{i}}}
\end{equation}

\noindent
where $\lambda$ is a Lagrangian multiplier that can be found by a bi-sectioning algorithm \cite{Bends,Sigmund1994}.

The sensitivity of the objective function is found as

\begin{equation}
\label{opt3}
   \frac{\partial p}{\partial \chi_{i}} = -\kappa (\chi_{i})^{\kappa-1}  v_{i}^{T}k_{i} v_{i}.
\end{equation}

To ensure existence of unique solution and mesh independency for the optimization problem, a filtering method \cite{Sigmund1994} is used to update element sensitivities as

\begin{equation}
\label{opt4}
   \frac{\hat{\partial p}}{\partial \chi_{i}} = \frac{1}{\sigma_{i}\sum^{N}_{j=1}D_{j}}\sigma_{i}\sum^{N}_{j=1}D_{j}\chi_{j}\frac{\partial p}{\partial \chi_{j}} 
\end{equation}

\noindent
where $D_{j}$ is the convolution operator which is

\begin{equation}
\label{opt5}
    D_{j} = r_{min} -\textrm{dist}(i,j)
\end{equation}

\noindent
where $j=\{ j \in N \mid \textrm{dist}(i,j) \leq r_{min}; i=1,...,N \}$ and operator dist($i,j$) is defined as the distance between centers of elements $i$ and $j$. 
Therefore, the convolution operator is zero outside the filtering area ($r_{min}$) and decays linearly inside the filter area. 
The modified sensitivities are used in the Optimality Criteria update.
$\widehat{\bm{\sigma}}$ depends on current stimulations, $\textbf{I}$.
This means that for each $I_{ij}$, there is a $\widehat{\bm{\sigma}}$.
$\widehat{\bm{\sigma}}$ is the optimum background electrical conductivity of $\Omega$ which may provide the highest $\Lambda$ for each $I_{ij}$.

For each ERT domain with current stimulations of $\textbf{I}= \{I_{ij} \}$, all of $\{\widehat{\bm{\sigma}}_{ij}\}$ should be found. 
The final optimum electrical conductivity background, $\widehat{\bm{\sigma}}_{\Lambda}$, is a nonlinear function of all $\{\widehat{\bm{\sigma}}_{ij}\} $.
Finally, the $\widehat{\bm{\sigma}}_{\Lambda}$ is found by maximizing $\Lambda$ over all $\textbf{I}= \{I_{ij} \}$ (Equation \ref{BBE1}).

\begin{eqnarray}
\label{BBE1}
\widehat{\bm{\sigma}}_{\Lambda} = \arg\min_{\bm{\sigma}} \Big\{ \| \textbf{V}_{ij}^{T}\textbf{K}_{ij}\textbf{V}_{ij}\|_{1/2}\Big\} \nonumber
\\
\textrm{which is subjected to:} \nonumber
\\
\frac{\bm{\phi}(\sigma_{\Lambda})}{\bm{\phi}_{0}}=\zeta 
\\
\textbf{K}_{ij} \textrm{V}_{ij}= \textrm{f}_{ij} \nonumber
\\
\bm{\sigma} = \bm{\chi} \otimes \bm{\sigma}_{0} \nonumber
\\
0.01 \leq \bm{\sigma} \leq 1. \nonumber
\end{eqnarray}

\noindent
where $\textbf{K}_{ij}$, $\textbf{V}_{ij}$ and $\textbf{f}_{ij}$ are stacked global stiffness, potentials and current matrices for all current stimulations pattern of  $\textbf{I}= \{I_{ij} \}$.
The optimization problem similarly is solved using standard Optimality Criteria method.
$\widehat{\bm{\sigma}}_{\Lambda}$ uses maximum amount of electrical power to transfer the injections for all current stimulation patterns which maximize the $\Lambda$ distribution over the sensor ($\widehat{\Lambda}$).

\section{ERT Inverse Problem}
 \label{IP}

In order to study whether distinguishability enhancement can provide reconstructability enhancement, difference imaging reconstructions are provided in this paper.
In difference imaging, we denote the conductivity distribution difference between before and after change by $\delta \sigma \in \textbf{R}^{N}$.
In difference imaging, an approximate, linearized observation model for the difference potential measurement data $\delta V \in \textbf{R}^{M}$ is written as

\begin{equation}
\label{est}
\delta V \approx J (\delta \sigma) + \delta n,
\end{equation}

\noindent
where $J$ is the Jacobian matrix of the mapping between spatially discretized conductivity changes $\delta \sigma$ and $\delta V$, and $\delta n \in \textbf{R}^{M}$ is the measurement noise.
Here, $M$ denotes the total number of potential measurements.
Given the above model, the objective in difference imaging is to estimate $\delta \sigma$ based on the $\delta V$, which leads to a minimization problem

\begin{equation}
\label{solution2}
\delta \sigma = \arg\min_{\delta \sigma} \Big\{ \| L_{\delta n}(\delta V - J \delta \sigma) \|^2 + P_{\delta \sigma}(\delta \sigma) \Big\},
\end{equation}

\noindent
where $P_{\delta \sigma}(\delta \sigma)$ is TV regularization functional related to the change of conductivity \cite{Rudin}.
The Cholesky factor of the noise precision matrix $L_{\delta n}$ is defined as $L_{\delta n}^{T}L_{\delta n}=\Gamma_{\delta n}^{-1}$, where $\Gamma_{\delta n}$, is the covariance of the noise.

Here, the regularization functional  $P_{\delta \sigma}(\delta \sigma)$ is selected as

\begin{equation}
\label{tv}
   P_{\delta \sigma}(\delta \sigma) = \eta \sum_{k=1}^{N} \sqrt{\| (\nabla \sigma) |_{\Omega_{k}} \|^2 + \Theta}
\end{equation}
\noindent
which is a differentiable approximation of the isotropic TV functional \cite{Rudin}.
Here, $(\nabla \sigma) |_{\Omega_{k}}$ is the gradient of the $\sigma$ at element $\Omega_{k}$ in the FEM mesh.
The choice of regularizing functional in Equation (\ref{solution2}) depends on the application.
In the present work, we use a TV regularization since the local anomaly change results in sharp boundary $\sigma$ change.
The regularization parameter $\eta$ is a weighting parameter which is chosen $7\times10^{-3}$ for all ERT reconstruction in this paper. 
$\Theta > 0$ is a small parameter that ensure TV is differentiable; $\Theta$ is chosen as $5\times10^{-8}$.
In order to provide ERT reconstructions, the simulated experimental measurements are provided first with added $0.1 \%$ normally distributed noise and with different forward model meshing from ERT difference imaging reconstructions.

\section{Results and discussion}
\label{RD}

Section \ref{TOP} described the approach to optimize electrical conductivity distribution.
Results of the optimization problem clearly depend on several factors.
Some of them were emphasized earlier in section \ref{ERT1} and illustrated by Figure 1.
For purpose of topology optimization, the new factor, $\zeta$, should be defined as electrical conductivity fraction of optimized domain to initial domain.
$\zeta$ certainly plays an important role in topology optimization.
The optimization problem above can be solved for different $\zeta$ values which result in different topology optimization solutions.
Therefore, for main part of this section we focus mainly on $\zeta = 0.6$.
$\zeta = 0.6$ means, topology optimization will result in an optimal design with $60 \%$ of electrical conductivity distributions of uniform background conductivity of $\sigma =1$ over the domain.
But finally, we compare results for multiple $\zeta$ designs for further discussion.

\begin{figure*}[ht!]
    \centering
\subfloat{{\includegraphics[width=3.3cm]{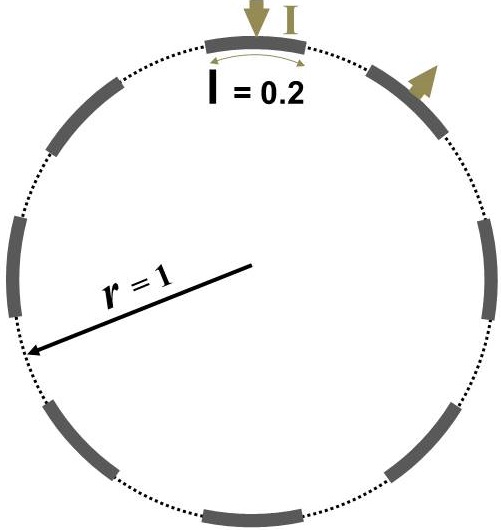} }}
    \quad
\subfloat{{\includegraphics[width=3.3cm]{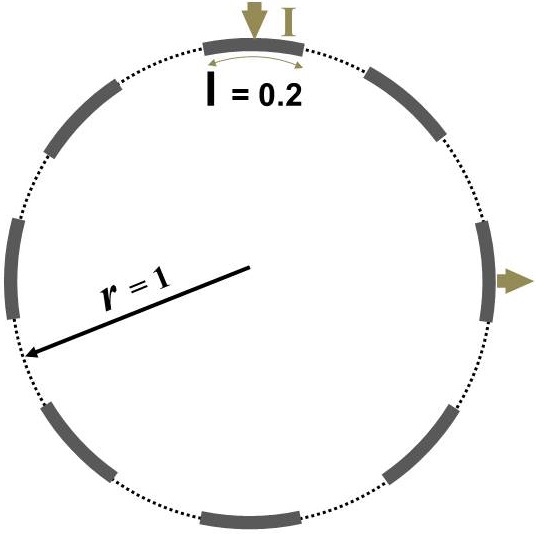} }}
    \quad
\subfloat{{\includegraphics[width=3.3cm]{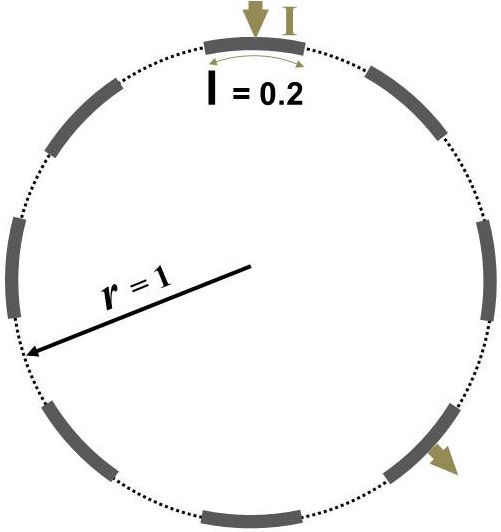} }}
    \quad
\subfloat{{\includegraphics[width=3.5cm]{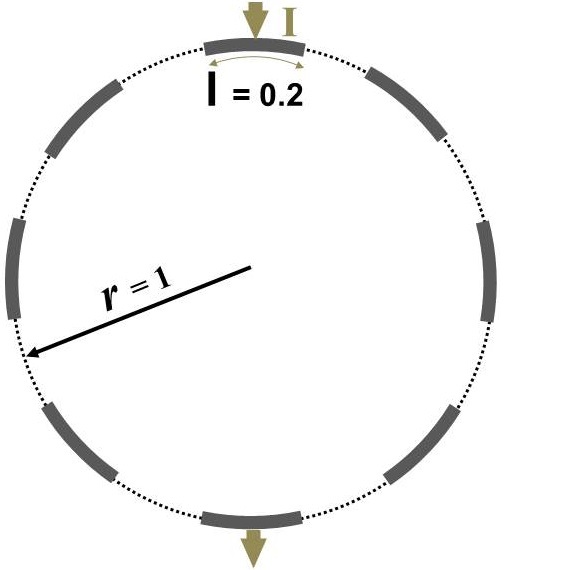} }}
 \\
\subfloat{{\includegraphics[width=3.1cm]{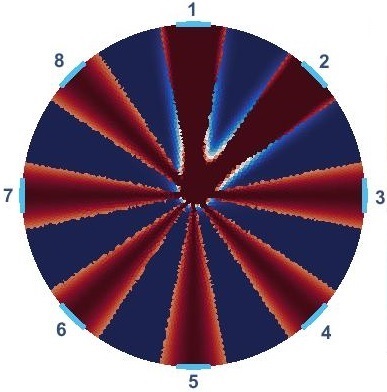} }}
    \quad
\subfloat{{\includegraphics[width=3.1cm]{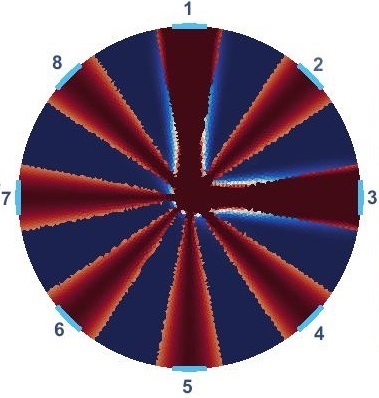} }}
    \quad
\subfloat{{\includegraphics[width=3.3cm]{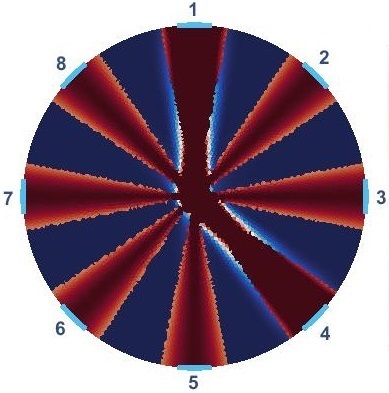} }}
    \quad
\subfloat{{\includegraphics[width=4cm]{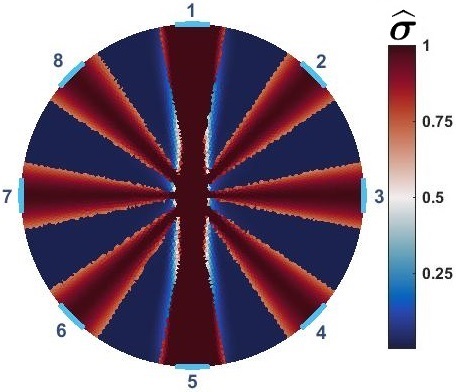} }}
 \caption{The first 4 current stimulations and corresponding $\widehat{\bm{\sigma}}$ distributions.}
 \label{Top1}
\end{figure*}

The domain shown in Figure 1 consists of a total of 56 patterns of current injections ($\textbf{I}_{ij}$).
Current injection is one of the prominent variables in optimization of background conductivity because of its effect on electrical potentials at electrodes.
First, we solve 56 individual optimization problems (Equation \ref{BBE1}) to obtain $\widehat{\bm{\sigma}}$ for each current injections.
Due to the symmetry, first 4 estimated $\widehat{\bm{\sigma}}$ corresponds to first set of current injections suffice to represent all $\widehat{\bm{\sigma}}_{ij}$ corresponding to $\textbf{I}_{ij}$.
The rest of 52 solutions are replications.
Figure \ref{Top1} shows the first four current stimulations and corresponding $\widehat{\bm{\sigma}}$ distributions with $\zeta = 0.6$.
Since $\widehat{\bm{\sigma}}$ stands for the background conductivity for maximum compliance inside the domain, it maximizes electrical potential of electrodes $\|\textbf{U}(\bm{\sigma}_{1},\textbf{I})\|$ which in-turn maximizes $\Lambda$.

$\widehat{\bm{\sigma}}$ results presented by Figure \ref{Top1} follow similar patterns and suggest periodic behavior in angular axis with higher conductivity closer to electrodes.
Their differences come from different $\textbf{I}_{ij}$.
Despite this difference, in all of them, the optimized background conductivity increases current density distribution at center of the domain.
In circular domain, center has the lowest distinguishability \cite{Rashetnia}.
The optimized $\widehat{\bm{\sigma}}$ provides a more uniform current density distribution specially at the center.
This enables ERT to provide more useful information at electrodes which means eventually higher $\Lambda$.
The next step is to investigate whether the optimized $\widehat{\bm{\sigma}}$  provides a better distinguishability for the corresponds current stimulations $I_{ij}$ than the uniform $\bm{\sigma}_{0}$.

\begin{figure*}[ht!]%
    \centering
\subfloat{{\includegraphics[width=3.1cm]{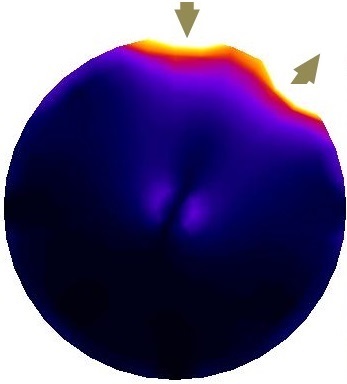} }}%
\quad  	
\subfloat{{\includegraphics[width=3.3cm]{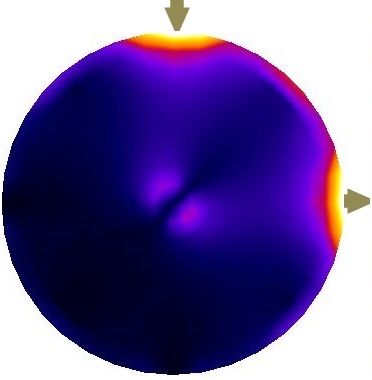} }}%
\quad
\subfloat{{\includegraphics[width=3.1cm]{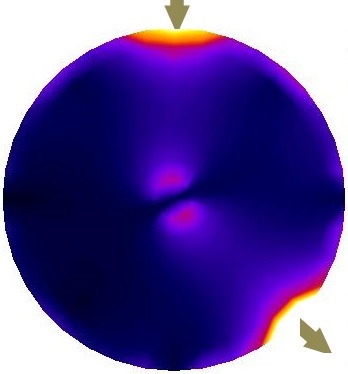} }}%
\quad
\subfloat{{\includegraphics[width=3.5cm]{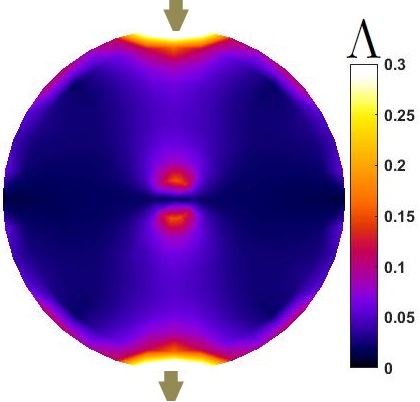} }}%
\\
\subfloat{{\includegraphics[width=3.1cm]{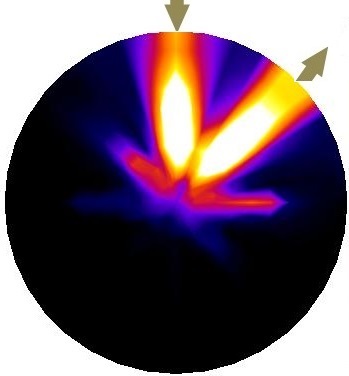} }}%
\quad
\subfloat{{\includegraphics[width=3.3cm]{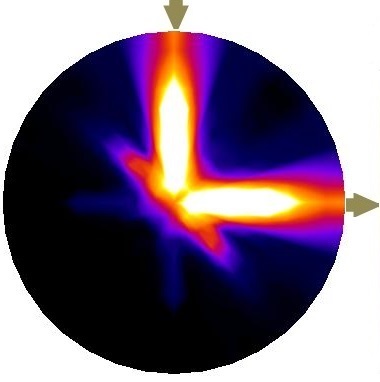} }}%
\quad
\subfloat{{\includegraphics[width=3.1cm]{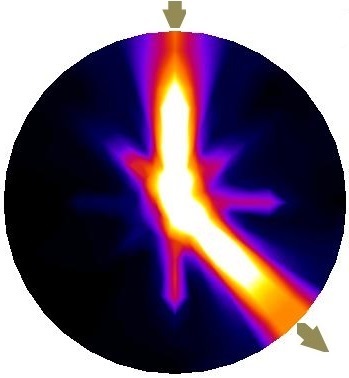} }}%
\quad
\subfloat{{\includegraphics[width=3.5cm]{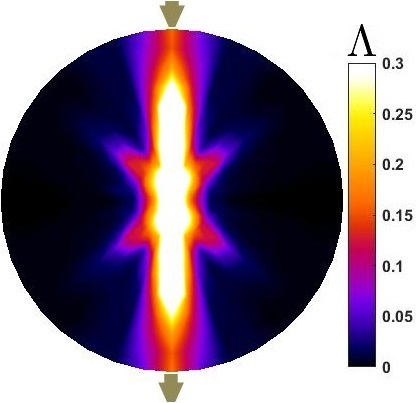} }}%
 \caption{Distinguishability distribution, $\Lambda$, over $\Phi$. The top row illustrates $\bm{\Lambda}_{0}$ over uniform $\bm{\sigma}_{0} = 1$ domain. The bottom row illustrates $\widehat{\bm{\Lambda}}$ over topological optimized domain ,$\widehat{\bm{\sigma}} \in \Omega$.}
 \label{RD2}%
\end{figure*}

For this purpose, The anomaly $\Psi$ ($\Psi \subset \Phi \subset \Omega$) is moved inside $\Phi$ for all $\widehat{\bm{\sigma}}$ and $\bm{\sigma}_{0}=1$, and $\Lambda$ is measured (shown in Figure 1b).
Figure \ref{RD2} compares $\widehat{\bm{\Lambda}}$ and $\bm{\Lambda}_{0}$ distributions for all four current injection scenarios.
In all four current injection scenarios, $\widehat{\bm{\sigma}}$ shows higher overall distinguishability ($\|\widehat{\bm{\Lambda}}\|>\|\bm{\Lambda}_{0}\|$).
$\widehat{\bm{\sigma}}$ provides maximum distinguishability at center and areas far from the boundaries for all the cases.
This in fact is an advantage since ERT based sensors suffer from asymptotic loss of distinguishability with increase of size because distance of the interior of sensor increased from electrodes \cite{garde2016depth,garde2017}.
Results suggest that maximizing compliance matrix of the sensors improve distinguishability of the areas far from electrodes.

\begin{figure*}[ht!]
    \centering
    \subfloat{{\includegraphics[width=5cm]{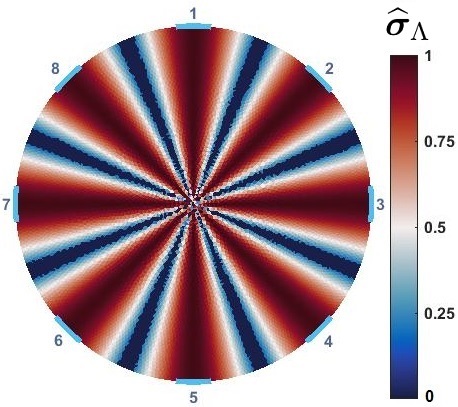} }}%
    \caption{Topological optimized domain ,$\widehat{\bm{\sigma}}_{\Lambda} \in \Omega$ for all current stimulation patterns ($\zeta = 0.6$).}   
    \label{RD5}
\end{figure*}

However, Figure \ref{RD2} shows $\|\widehat{\bm{\Lambda}}\|>\|\bm{\Lambda}_{0}\|$; areas of $\widehat{\bm{\sigma}}$ with minimum 0.001 values of conductivity gain very low distinguishability which may become blinds spots.
This may make $\bm{\sigma}_{0}$ more desirable than $\widehat{\bm{\sigma}}$ due to this issue.
But, proposed results are from single current stimulations. 
Therefore, having multiple current stimulation patterns may fix this issue.
Then, next step is to estimate $\widehat{\bm{\sigma}}_{\Lambda}$ corresponds to all current stimulations for case of $\zeta = 0.6$ which is shown by Figure \ref{RD5}.

$\widehat{\bm{\sigma}}_{\Lambda}$ presents the optimum distribution of conductivity in the domain which maximizes electrical power function inside the sensor while improve performance and distinguishability of the sensor. 
$\widehat{\bm{\sigma}}_{\Lambda}$ is shown as a symmetric angular distribution of conductivity patterns.
Electrical conductivity is higher angularly closer to the electrodes, which means these regions have more contribution in power transfer function.
This pattern makes electrical current travel the most optimal path for each $I_{ij}$.

\begin{figure*}[ht!]%
    \centering
\subfloat[]{{\includegraphics[width=4.1cm]{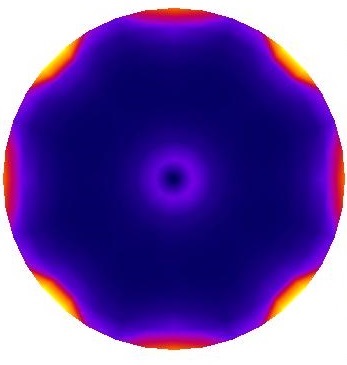} }}%
\quad
\subfloat[]{{\includegraphics[width=5cm]{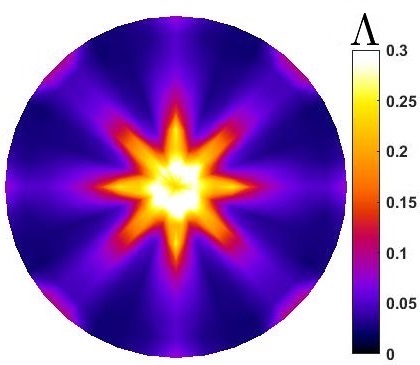} }}%
 \caption{a) $\bm{\Lambda}_{0}$ distribution over uniform $\sigma = 1$ domains. b) $\widehat{\bm{\Lambda}}$ distribution over topological optimized $\widehat{\bm{\sigma}}_{\Lambda}$.}
 \label{RD6}%
\end{figure*}

Figure \ref{RD6} shows the corresponds distinguishability distributions of  $\bm{\Lambda}_{0}$ and $\widehat{\bm{\Lambda}}$.
$\widehat{\bm{\sigma}}_{\Lambda}$ improves distinguishability from uniform background significantly.
It specifically improves the $\Lambda$ values in lowest distinguishability areas which are over center region.
Therefore, using full current stimulation patterns, using optimized sensors ($\Lambda$) provide more "visibility".
This suggests that, with provided dimensions, number of electrodes and size of sensor, one way of improving the distinguishability of the sensor is engineered background electrical conductivity.
This is an applicable approach to improve ERT sensor for application of large area sensing \cite{Rashetnia2018,Rashetnia}.
This can show a possible application of these sensors for the cases that sensors will be used with higher probability of changes at certain regions of interest.
Further, proposed results prove that, using $\widehat{\bm{\sigma}}_{\Lambda}$ with several current stimulation patterns provides even better distinguishability distribution.

\begin{figure*}[ht!]%
    \centering
\subfloat[$\zeta=0.2$]{{\includegraphics[width=2.3cm]{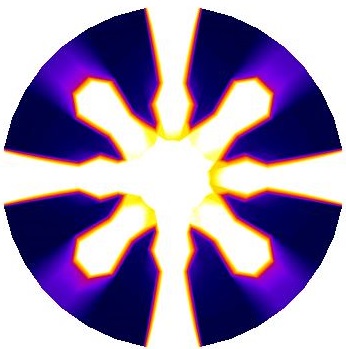} }}%
\quad
\subfloat[$\zeta=0.4$]{{\includegraphics[width=2.3cm]{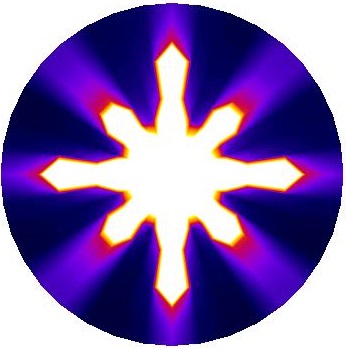} }}%
\quad
\subfloat[$\zeta=0.5$]{{\includegraphics[width=2.3cm]{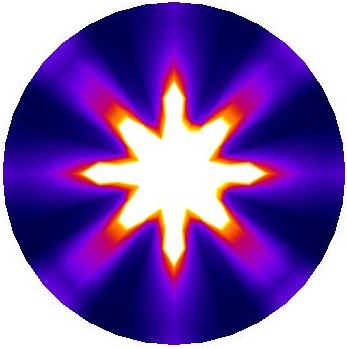} }}%
\quad
\subfloat[$\zeta=0.6$]{{\includegraphics[width=2.3cm]{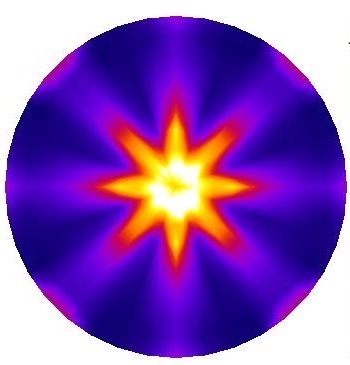} }}%
\quad
\subfloat[$\zeta=0.7$]{{\includegraphics[width=2.3cm]{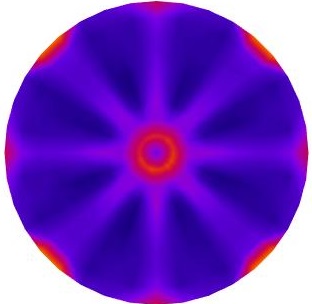} }}%
\quad
\subfloat[$\zeta=1$]{{\includegraphics[width=2.8cm]{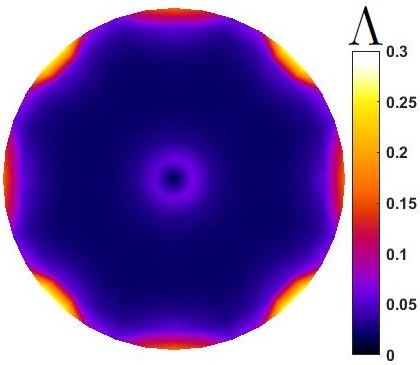} }}%
 \caption{$\widehat{\bm{\Lambda}}$ distribution over topological optimized $\widehat{\bm{\sigma}}_{\Lambda}$ with different $\zeta$ parameter values.}
 \label{RD7}%
\end{figure*}

$\zeta$ is a parameter in topology optimization which should be fixed as $\zeta \leq 1$.
$\zeta = 1$ means electrical conductivity distribution equal to whole fraction of uniform background distribution of $\sigma_{0} = 1$.
The constrained condition of $\zeta$ provides the electrical conductivity fraction of the optimal background domain.
Therefore, value of $\zeta$ can affect the topology optimized answer and corresponding $\widehat{\bm{\Lambda}}$.
Figure \ref{RD7} compares $\widehat{\bm{\Lambda}}$ distributions for $\zeta$ equal to 1, 0.7, 0.6, 0.5, 0.4, and 0.2.
$\zeta = 1$ means the uniform $\bm{\sigma}_{0}$.
Higher $\zeta$ results in higher distinguishability over central regions.
Although, in lower ranges of $\zeta \leq 0.5$ distinguishability increased at central parts but reduced significantly at the other regions.
Comparing Figure \ref{RD6} results, it can be shown that $0.5 \leq \zeta \leq 1$ improves distinguishability all around the domain.

\begin{figure*}[ht!]
    \centering
\subfloat{{\includegraphics[width=3.0cm]{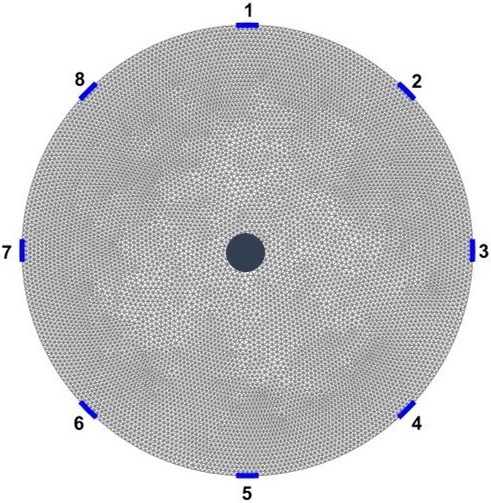} }}
    \quad
\subfloat{{\includegraphics[width=3.0cm]{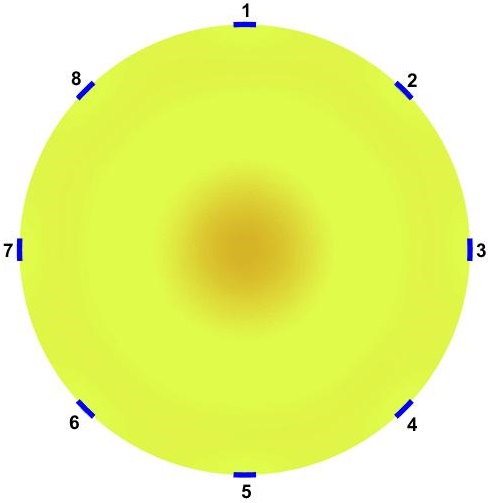} }}
    \quad
\subfloat{{\includegraphics[width=3.5cm]{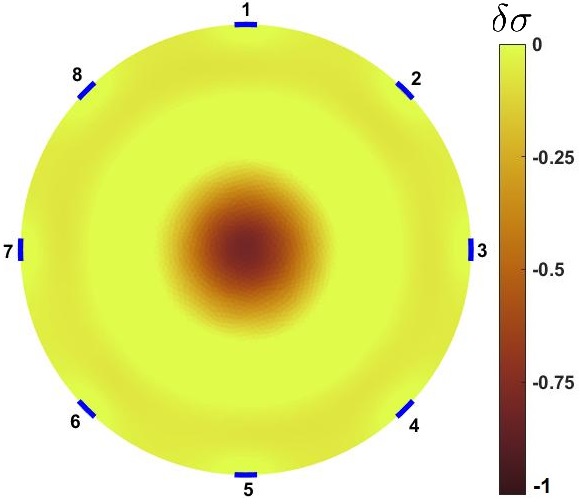} }}
 \\
\subfloat{{\includegraphics[width=3.0cm]{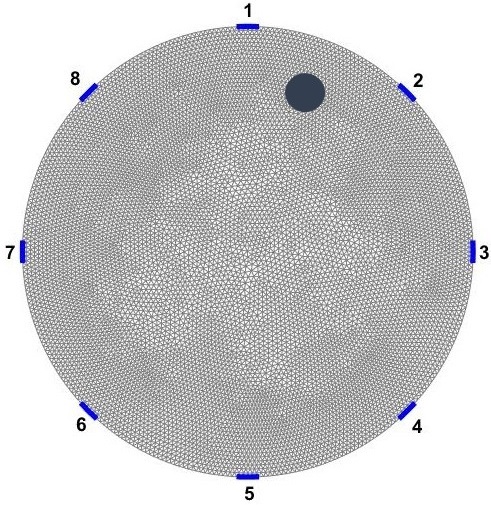} }}
    \quad
\subfloat{{\includegraphics[width=3.0cm]{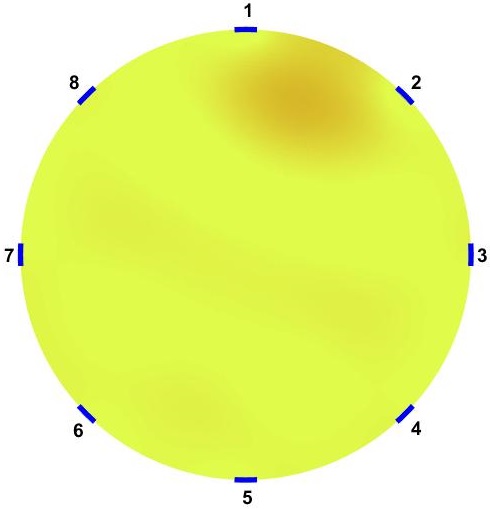} }}
    \quad
\subfloat{{\includegraphics[width=3.5cm]{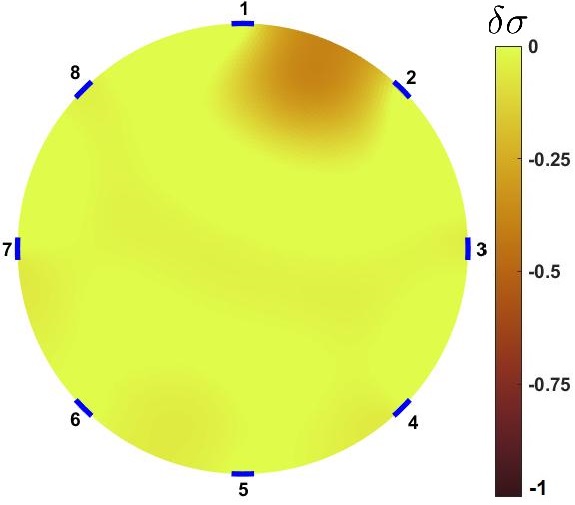} }}
\\
\subfloat{{\includegraphics[width=3.0cm]{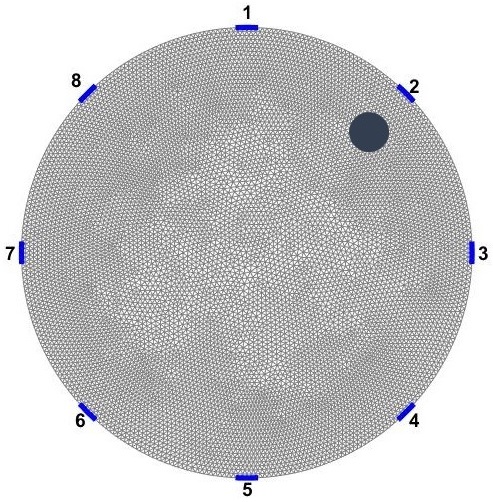} }}
    \quad
\subfloat{{\includegraphics[width=3.0cm]{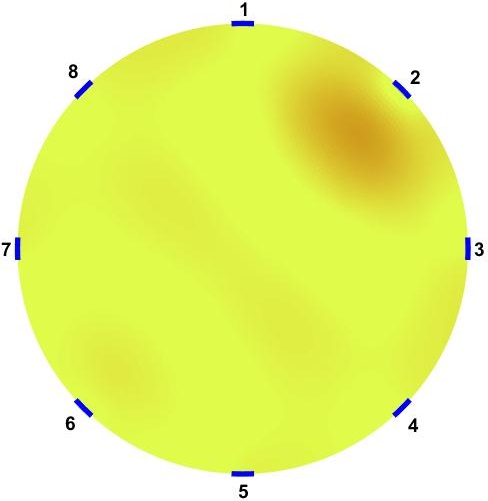} }}
    \quad
\subfloat{{\includegraphics[width=3.5cm]{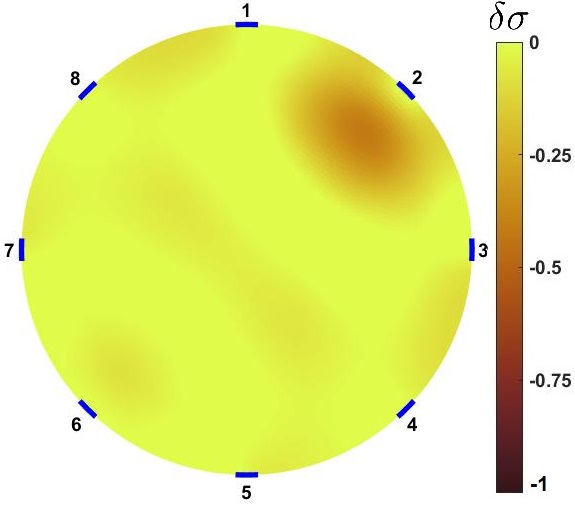} }}
 \caption{first column: circular anomaly with $\sigma =0.001$ are placed at center, over the edge between electrodes 1 and 2 and in front of electrode 2; second column: ERT reconstructions over uniform electrical conductivity of $\sigma =1$; third column: $\delta \widehat{\bm{ \sigma}}_{\Lambda}$ ERT reconstructions over $\widehat{\bm{\sigma}}_{\Lambda}$ distribution with $\zeta=0.6$.}
 \label{R}
\end{figure*}

In order to study whether distinguishability enhancement can provide a better reconstructability, three difference imaging reconstructions are provided.
The first column of Figure \ref{R} presents two circular anomalies with 0.001 $s$ conductivity and $0.05$ radius in center and the edge of sensor.
The anomalies were placed at center, close to edge  between electrodes 1 and 2, and close to electrode 2 because these three spots have highest and lowest distinguishability values in optimized sensors (Figure \ref{RD7}).
$\bm{\delta \sigma}$ are reconstructed over uniform background conductivity of $\sigma =1$ (second column of Figure \ref{R}). $\delta \widehat{\bm{ \sigma}}_{\Lambda}$ also reconstructed for optimized background conductivity $\widehat{\bm{\sigma}}_{\Lambda}$ with $\zeta=0.6$ (third column of Figure \ref{R}), respectively.
For all three cases, anomaly detected sharper and closer to real changes in $\widehat{\bm{\sigma}}_{\Lambda}$ than uniform background.
These result can support the idea that distinguishability enhancement can provide reconstructability enhancement as well.
Also, comparing both recosntruction cases suggests that using $\widehat{\bm{\sigma}}_{\Lambda}$ instead of uniform background does not reduce reconstruction ability of ERT over edges and it does not result in blind spots.

\section{Conclusion}

In this paper, we proposed an optimization approach for obtaining an optimal background conductivity distribution which results in highest distinguishability distribution for ERT based sensors.
The proposed approach was evaluated using ERT simulations over a circular sensor with 8 electrodes.
It has been shown that this approach provides the background conductivity distribution with higher distinguishability ($\| \Lambda \|$).
Also, the $\zeta$ parameter as a designing target conductivity volume can affect the optimized results. 

In proposed approach, the ERT domain topologically was optimized to maximize compliance matrix over the domain while kept the performance higher than uniform background domain.
The higher performance is achieved because of better current density distribution over the domain which means that we measure more current density changes over the domain using optimized background conductivity.
Also, we guide the current flow into the center of the domain which is furthest point from electrodes.
In regular ERT domains, distinguishability decreased asymptotically as a function of distance to the electrodes.
Therefore, proposed design improves distinguishability of the areas with least distinguishability expectations.

Results in Figure \ref{RD6} suggests that with using less material and using optimum design we can improve distinguishability specially at most critical regions or regions of interest.
The region of interest can be chosen as regions far from boundaries, ERT provides lowest distinguishability values, or any susceptible area of the domain.
This approach, can be used for application which region of interest is available, which the customized design improves distinguishability over regions of interest. 
Using optimized ERT domain, distinguishability is improved at region of interest significantly and potentially we can achieve better reconstructability.
It also enables ERT to be used for applications with higher amount of noises and using ERT experimentations with less of the experimental precision of measurements.
Results in Figure \ref{RD6} shows effect of $\zeta$ values on final results.

Finally, ERT reconstructions were presented to investigate whether distinguishability enhancement can provide reconstructability enhancement. 
Difference imaging with TV regularization applied to both uniform and optimized background conductivity, and results suggest that optimized background conductivity may improve reconstructability.

\section*{References}

\end{document}